\documentclass[prd,twocolumn,groupedaddress,superscriptaddress,lengthcheck]{revtex4}
\usepackage[a4paper, total={7in, 9in}]{geometry}
\usepackage{times}
\usepackage{mathptm}
\usepackage{amsmath}

\usepackage{graphicx}  
\usepackage{caption}
\usepackage{subfigure}
\usepackage{multirow}
\usepackage{float}
\usepackage{fancyhdr}
\usepackage{parskip}
\usepackage[T1]{fontenc}
\usepackage{dcolumn}   

\usepackage{bm}        
\usepackage{amsfonts}  
\usepackage{amsmath}   
\usepackage{amssymb}   
\usepackage{hyperref}
\hypersetup{
    colorlinks=true,
    linkcolor=blue,
    citecolor=blue,
    filecolor=blue,      
    urlcolor=blue,
}

\begin{document}

\title{Detectability of subsolar mass neutron stars through a template bank search}

\author{Ananya Bandopadhyay}\email{abandopa@syr.edu}
\affiliation{Department of Physics, Syracuse University, Syracuse, NY 13244, USA}

\author{Brendan Reed}\email{reedbr@iu.edu}
\affiliation{Department of Astronomy, Indiana University,
                  Bloomington, IN 47405, USA}
\affiliation{Center for Exploration of Energy and Matter and
                  Department of Physics, Indiana University,
                  Bloomington, IN 47405, USA}  
                  
\author{Surendra Padamata}\email{ssp5361@psu.edu}
\affiliation{Institute for Gravitation and the Cosmos, The Pennsylvania State University, University Park, PA 16802, USA}
\affiliation{Department of Physics, The Pennsylvania State University, University Park, PA 16802, USA}

\author{Erick Leon}
\affiliation{Department of Physics, Syracuse University, Syracuse, NY 13244, USA}

\author{C. J. Horowitz}\email{horowit@indiana.edu}
\affiliation{Center for Exploration of Energy and Matter and
                  Department of Physics, Indiana University,
                  Bloomington, IN 47405, USA}
           
\author{Duncan A.~Brown}
\affiliation{Department of Physics, Syracuse University, Syracuse, NY 13244, USA}

\author{David Radice}\thanks{Alfred P.~Sloan Fellow}
\affiliation{Institute for Gravitation and the Cosmos, The Pennsylvania State University, University Park, PA 16802, USA}
\affiliation{Department of Physics, The Pennsylvania State University, University Park, PA 16802, USA}
\affiliation{Department of Astronomy \& Astrophysics, The Pennsylvania State University, University Park, PA 16802, USA}

\author{F.~J. Fattoyev}\email{ffattoyev01@manhattan.edu}
\affiliation{Department of Physics \& Astronomy, Manhattan College, Riverdale, NY 10471, USA}

\author{J. Piekarewicz}\email{jpiekarewicz@fsu.edu}
\affiliation{Department of Physics, Florida State University,
               Tallahassee, FL 32306, USA}

\date{\today}

\begin{abstract}
    We study the detectability of gravitational-wave signals from subsolar-mass binary neutron star systems by the current generation of ground-based gravitational-wave detectors. We find that finite size effects from large tidal deformabilities of the neutron stars and lower merger frequencies can significantly impact the sensitivity of the detectors to these sources. By simulating a matched-filter based search using injected binary neutron star signals with tidal deformabilities derived from physically motivated equations of state, we calculate the reduction in sensitivity of the detectors. We conclude that the loss in sensitive volume can be as high as $78.4 \%$ for an equal mass binary system of chirp mass $0.17 \, \textrm{M}_{\odot}$, in a search conducted using binary black hole template banks.  We use this loss in sensitive volume, in combination with the results from the search for subsolar-mass binaries conducted on data collected by the LIGO-Virgo observatories during their first three observing runs, to obtain a conservative upper limit on the merger rate of subsolar-mass binary neutron stars. Since the discovery of a low-mass neutron star would provide new insight into formation mechanisms of neutron stars and further constrain the equation of state of dense nuclear matter, our result merits a dedicated search for subsolar-mass binary neutron star signals. 
\end{abstract}

\maketitle

\section{Introduction} 

The first detection of a gravitational wave signal from a binary neutron star merger, GW170817, \cite{LIGOScientific:2017vwq} allowed the exploration of matter at high densities~\cite{LIGOScientific:2018cki,De:2018uhw,Radice:2018ozg,Raithel:2018ncd,Capano:2019eae}. The LIGO-Virgo network of second-generation gravitational wave observatories have observed neutron stars in four compact binary coalescence events. Two of these were binary neutron star mergers, and the other two neutron star black hole mergers \cite{LIGOScientific:2017vwq,LIGOScientific:2020aai,LIGOScientific:2021qlt}. Among these, matter was only detected  for the GW170817 event, which was accompanied by a multi-wavelength electromagnetic counterpart that was observed across the entire electromagnetic spectrum~\cite{LIGOScientific:2017ync}. The masses of the neutron stars observed in these events are greater than $1\,\rm M_{\odot}$. Separately from these, gravitational wave events involving subsolar-mass compact objects provide another important discovery space. The observation of even a single subsolar-mass event could be a revolutionary discovery, and can imply the existence of qualitatively new kinds of exotic objects, since there are no known astrophysical processes involving subsolar-mass compact objects that produce gravitational wave signals detectable by the LIGO-Virgo detectors~\cite{Sotani:2013dga}. \\
The recent claim of the observation of a $0.77 \, \rm M_{\odot}$ neutron star within a supernova remnant \cite{Doroshenko2022} suggests that subsolar-mass neutron stars could be an important class of sources for gravitational wave detectors. The minimum theoretical mass of a neutron star constructed from a cold, dense equation of state is $\approx 0.1\, \textrm{M}_{\odot}$ \cite{ShapiroTeukolsky}. However, the supernova mechanism likely imposes a minimum mass larger than this \cite{Lattimer:2000nx}. This constraint provides a lower bound on the allowed masses of neutron stars, but it depends on supernova physics that is not yet fully understood. Very low mass neutron stars with masses down to about $0.1 \, \textrm{M}_\odot$ are still gravitationally bound and could be formed through speculative channels, such as the fragmentation of a neutron star in a collision~\cite{Popov:2006ki}. The $0.2 \, \rm M_\odot$ object discovered by self-lensing in the KIC 8145411 system could be an example of such a low-mass neutron star, formed in an unexpected way~\cite{Masuda_2019}. As pointed out by Silva {\it et al.} \cite{Silva:2016myw}, there are various observational and theoretical reasons why low-mass neutron stars are an interesting target for gravitational-wave detectors. 

The cores of neutron stars can reach densities up to 5-6 times the nuclear saturation density $n_0 = 0.16 \, \rm fm^{-3}.$ The phase of gravitational wave signals emitted by inspiralling binary neutron star systems carries information about tidal effects on neutron star matter, which is strongly dependent on the nuclear equation of state \cite{Hinderer:2009ca,Chatziioannou:2020pqz}. Extracting information about the tidal polarizability of neutron stars from gravitational wave signals allows us to probe the dense low temperature region of the phase space of nuclear matter. This region is inaccessible through the existing theoretical tools, like chiral Effective  Field Theory or perturbative Quantum Chromodynamics \cite{Drischler:2021bup}, or through laboratory experiments \cite{PREX2,CREX} which probe the low density regime. The constraints on the tidal deformability of neutron stars from GW170817  \cite{LIGOScientific:2018cki,De:2018uhw,Radice:2018ozg,Raithel:2018ncd,Capano:2019eae}, and those on neutron star radius, obtained from the NICER X-ray telescope~\cite{Riley:2019yda, Miller:2019cac,Miller:2021qha}, have ruled out a large fraction of the equations of state which were previously considered feasible.

Measurements of the tidal deformability of neutron stars with masses less than $1\,\rm M_{\odot}$ can provide very interesting constraints on the structure of the crust of a neutron star. The central density of neutron stars with $M\lesssim 0.5 \, \textrm{M}_\odot$ is $n\sim 0.2  $fm$^{-3}$, with central densities near the minimum mass being as low as $n\sim 0.1  $fm$^{-3}$. As a result, these stars have a large crust compared to higher mass neutron stars, and their tidal deformabilities are largely insensitive to the equation of state of the core~\cite{Carrier:2003}. While the exact equation of state of the crust has been shown to not affect the tidal deformability parameter, $\Lambda$ (defined in Eq.~(\ref{eq:Lambda})), for a fiducial mass neutron star ($M=1.4 \, \rm M_\odot$) \cite{Piekarewicz:2019}, the behavior of $\Lambda$ shows a noticeable dependence on the crust near the minimum neutron star mass. Therefore, measurements of tidal deformability for a neutron star near $0.1 \, \rm M_\odot$ could help constrain the behavior of the neutron star crust directly. However, we are largely insensitive to the equation of state of the core because of the low densities of subsolar-mass neutron stars~\cite{Carrier:2003}.

Current searches for binary neutron stars are not designed to be sensitive to systems having component masses below $1\, \textrm{M}_{\odot}$, even though they retain some sensitivity in this regime. LIGO and Virgo have searched for gravitational wave signals from subsolar-mass black holes, with component masses lying in the range $0.2 - 1 \,  \textrm{M}_\odot$ \cite{LIGOScientific:2018glc,LIGOScientific:2019kan,LIGOScientific:2021job,LIGOScientific:2022hai}. These searches assume that the compact objects have a dimensionless tidal deformability $\Lambda_{1,2} = 0$, and the inspiral waveforms used as templates are terminated at their Nyquist frequency, which is 1024 Hz. Matter effects have been previously shown to impact searches for gravitational waves from binary neutron star systems in~\cite{Cullen:2017oaz}.

In this paper, we assess the impact of neglecting the tidal deformabilities and physical merger frequencies in a search for  binary neutron star systems with component masses $m_1,m_2 \in [0.2,1.0]\, \rm M_{\odot}$. We find that this can cause a non-trivial loss in detected event rate, due to the mismatch between the signal and template waveforms, in a matched-filter search. We use this result, and the result of the search for primordial black holes in the data collected by the LIGO and Virgo detectors during the O1 to O3 observing runs, reported in~\cite{Nitz:2022ltl}, to set an upper limit on the merger rate of subsolar-mass binary neutron stars.

The organization of the paper is as follows: In Sec.~\ref{section2} and Sec.~\ref{section3}, we explore the effects of large tidal deformabilities and lower merger frequencies resulting from large radii of subsolar-mass neutron stars, on the gravitational signals emitted by these sources. In Sec.~\ref{section4}, we analyze the detectability of these sources using binary black hole template banks, which are conventionally used during searches for subsolar-mass binaries, by simulating a search for injected subsolar-mass binary neutron star signals in the Advanced LIGO noise curve. We use the PyCBC toolkit~\cite{alex_nitz_2022_6912865} to perform the analysis presented in this section. In Sec.~\ref{section5} we estimate an upper limit on the merger rate of subsolar-mass binary neutron star systems by combining the reduction in sensitive volume of the detectors with the rate upper limit on subsolar-mass black hole binaries from a previous search, conducted on the data from the LIGO and Virgo observatories through the completion of their third observing run. Sec.~\ref{section6} includes a discussion of our main results and the potential astrophysical implications of observations of gravitational wave signals from binary neutron star mergers involving subsolar-mass component stars.

\section{Equation of State Models and the Tidal Deformability of Subsolar-Mass Neutron Stars} \label{section2}

 Neutron stars in binary systems are deformed by the tidal forces of their companion object, leading to deviations from spherical symmetry. The tidal deformability of a neutron star is defined as the constant of proportionality between the induced quadrupolar response, $\mathcal{Q}_{ij}$, of the star and the external perturbing field, $\mathcal{E}_{ij}$, that is,  $\mathcal{Q}_{ij} = -\lambda {\mathcal{E}_{ij}}$ \cite{Chatziioannou:2020pqz}. 

 \begin{figure*}[t]
\centering
\includegraphics[width=1\textwidth]{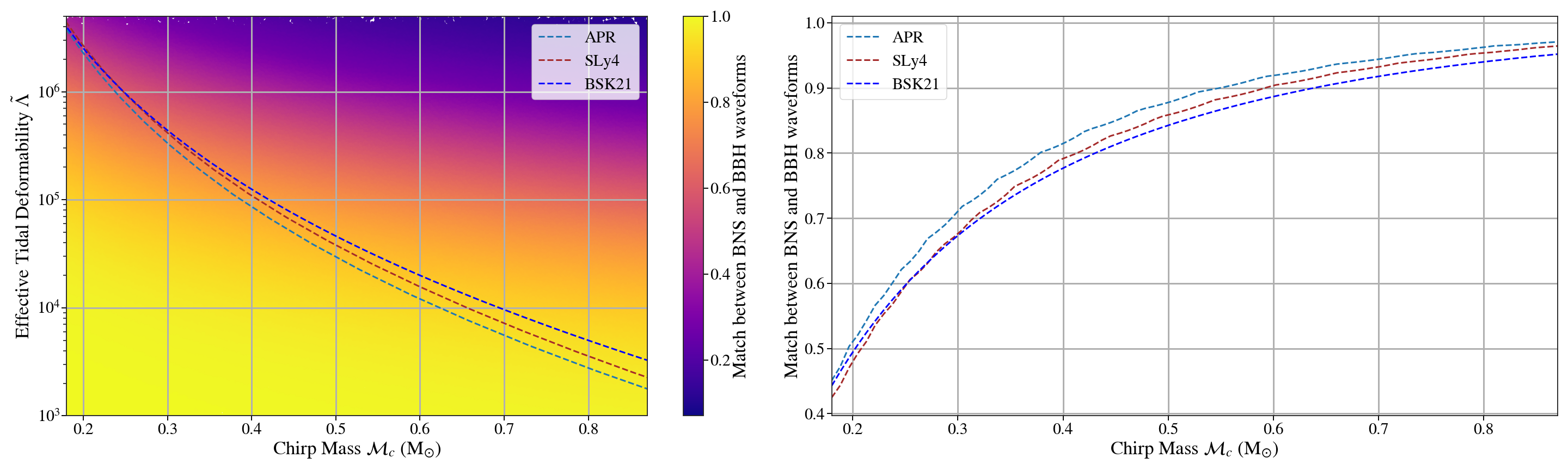}
\caption{Left panel shows the match between binary neutron star and binary black hole waveforms as a function of chirp mass $\mathcal{M}_C$ and effective tidal deformability $\tilde{\Lambda}$ of the binary system, for equal mass binaries. The binary neutron star signals are terminated at the frequency of gravitational wave emission at the innermost stable circular orbit for a point particle orbiting a Schwarzschild black hole. Low match values in the region bounded by the equation of state curves indicate that ignoring the tidal deformability of subsolar-mass neutron stars can reduce the sensitivity of LIGO searches to gravitational wave signals from these sources. Right panel shows the match between binary neutron star and binary black hole waveforms as a function of the chirp mass of the system, for equal mass binaries, and for three different neutron star equations of state. The lower frequency cutoff for match evaluation is 15 Hz.}
\label{bns_isco_overlap-mass-lambda}
\end{figure*}

The phase of gravitational waves emitted by inspiralling binary neutron star systems is sensitive to the nuclear equation of state through the effective tidal deformability parameter for the binary system. The tidal deformability parameter of a neutron star, in its dimensionless form, is given by~\cite{Hinderer:2009ca,Chatziioannou:2020pqz}
\begin{equation}
    \Lambda = \frac{\lambda}{(Gm/c^2)^5} = \frac{2}{3} k_2 \left( \frac{G m}{R c^2}\right)^{-5} 
    \label{eq:Lambda}
\end{equation}
Here, $k_2$ is the quadrupolar ($l=2$) tidal Love number, which characterizes the neutron star's response to an external tidal field, $R$ is its radius, and $m$ its mass~\cite{Damour:2009vw}. For neutron stars having comparable masses, a higher value of $\Lambda$ is indicative of a stiffer equation of state, and hence a more deformable structure. The leading order tidal correction to the phase of the gravitational waveform which occurs at 5 PN (post-Newtonian)  order is expressed in terms of the effective tidal deformability parameter, $\tilde{\Lambda}$ of the binary neutron star system, \cite{Chatziioannou:2020pqz,Flanagan:2007ix,Hinderer:2009ca} 
\begin{equation}
    \tilde{\Lambda} = \frac{16}{13} \frac{\left( m_1 + 12 m_2 \right) m_1^4 \Lambda_1 + \left( m_2 + 12 m_1 \right) m_2^4 \Lambda_2}{\left( m_1 + m_2\right)^5}
    \label{tilde-Lambda}
\end{equation}

where $m_1$, $m_2$ are the masses and $\Lambda_1$, $\Lambda_2$ are the dimensionless tidal deformabilities of the component neutron stars.\\
To account for the uncertainties in the nuclear equation of state at densities encountered in neutron stars, we assume three different equation of state models, namely the APR \cite{Akmal:1998cf}, SLy4 \cite{Douchin:2001sv} and BSk21 \cite{Pearson:2011zz,Pearson:2012hz,Potekhin:2013qqa} equations of state. These models all use a unified non-relativistic formalism and produce neutron stars which are consistent with both the maximum mass of known neutron stars \cite{NANOGrav:2019jur}, the measured values of neutron star radii from NICER, and $\Lambda_{1.4}$ from GW170817~\cite{LIGOScientific:2018cki,De:2018uhw,Radice:2018ozg,Raithel:2018ncd,Capano:2019eae,Riley:2019yda, Miller:2019cac,Miller:2021qha,Al-Mamun:2020vzu}. This set of equations of state is by no means comprehensive, but provides a range of possible neutron stars consistent with observational constraints. Using the equation of state models as inputs to the Tolman–Oppenheimer–Volkoff system of differential equations, and then following the approach discussed in \cite{Hinderer:2009ca}, we evaluate the tidal deformability of the neutron stars, which is then used to calculate the effective tidal deformability of the binary system using Eq.~(\ref{tilde-Lambda}). 

To demonstrate the effect of tidal deformabilities of subsolar-mass neutron stars on the gravitational wave signals emitted by inspiralling binary neutron star systems, we generate a set of waveforms for equal mass binary neutron stars, with chirp masses in the range $\mathcal{M}_C \in [0.18 ,0.87] \,  \textrm{M}_\odot$, and effective tidal deformabilities in the range $\tilde{\Lambda} \in [1000,5\times 10^6]$. We compute the match between each injected waveform and a binary black hole waveform having the same component masses, where the match quantifies the resemblance between the two. It is the overlap between the two waveforms, weighted by the Advanced LIGO noise power spectral density, maximized over the extrinsic paremeters of the waveform. The overlap integral for match calculation is evaluated from a lower frequency cutoff of 15 Hz, upto the Schwarzschild ISCO frequency. The plot on the left panel of Fig.~(\ref{bns_isco_overlap-mass-lambda}) shows the match as a function of chirp mass, $\mathcal{M}_C = \frac{(m_1 m_2)^{3/5}}{(m_1+m_2)^{1/5}}$, and effective tidal deformability $\tilde{\Lambda}$, of the binary neutron star systems. The region bounded by the equation of state curves overlaid on the figure approximately represents the feasible part of the parameter space for subsolar-mass binary neutron star systems. As can be seen in the figure, the tidal deformabilities of the subsolar-mass neutron stars are large, and the gravitational wave signals emitted by them can be significantly different from binary black hole signals, as is indicated by the low match values in the plot. This suggests that it is important to account for the tidal deformabilities of subsolar-mass neutron stars in order to accurately model the gravitational waveforms emitted by these sources. 

\section{Merger Frequencies of Subsolar-Mass Neutron Stars} \label{section3}

\begin{figure}[b]
    \centering
    \includegraphics[width=8cm]{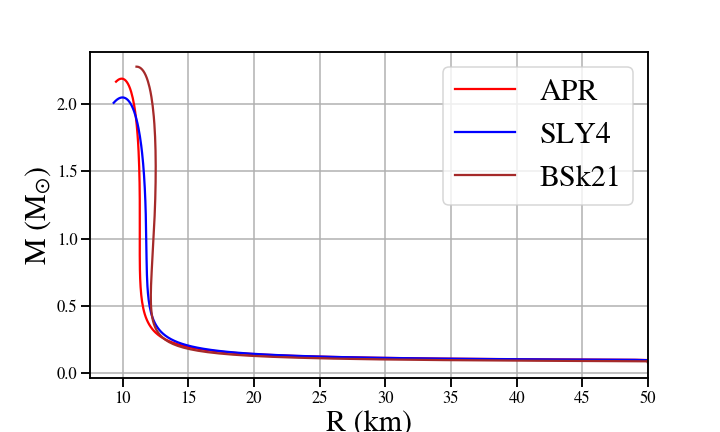}
    \caption{Comparison of mass-radius curves for neutron stars, obtained by integrating the Tolman-Oppenheimer-Volkoff system of differential equations for the three different equations of state being considered in this work. The large radii of the low mass neutron stars impact the merger frequency of their gravitational waveforms as they are tidally disrupted by the companion at lower frequencies than higher mass neutron stars.}
    \label{mass-radius-curves}
\end{figure}

In addition to assuming that the sources are binary black holes, the waveform models used in subsolar-mass searches also terminate the binary inspiral templates at the Nyquist frequency, 1024 Hz \cite{LIGOScientific:2018glc,LIGOScientific:2019kan,LIGOScientific:2021job}. This assumption is justifiable for binary black holes and more massive binary neutron star systems, since these are extremely compact objects. Subsolar-mass neutron stars, on the other hand, are significantly less compact, and have large radii, as can be seen in Fig.~(\ref{mass-radius-curves}). In the case of binaries in which one or both of the stars have a large radius and large tidal deformability, such as the low-mass binaries considered here, the merger is expected to take place shortly after Roche overflow of the secondary star \cite{Dietrich:2015pxa, Bernuzzi:2020txg}. As such, the point of Roche lobe overflow can be used as a conservative time at which to terminate an inspiral gravitational waveform model \cite{Shibata:2001ag, Marronetti:2003hx}. Here, we estimate the gravitational wave frequency corresponding to the mass-shedding limit $f_{\rm Roche}$ by performing a sequence of numerical simulations of binary neutron stars based on quasi-circular equilibrium approximation \cite{Bejger:2004zx}. \\

\begin{figure}[b]
    \centering
    \includegraphics[width=8cm]{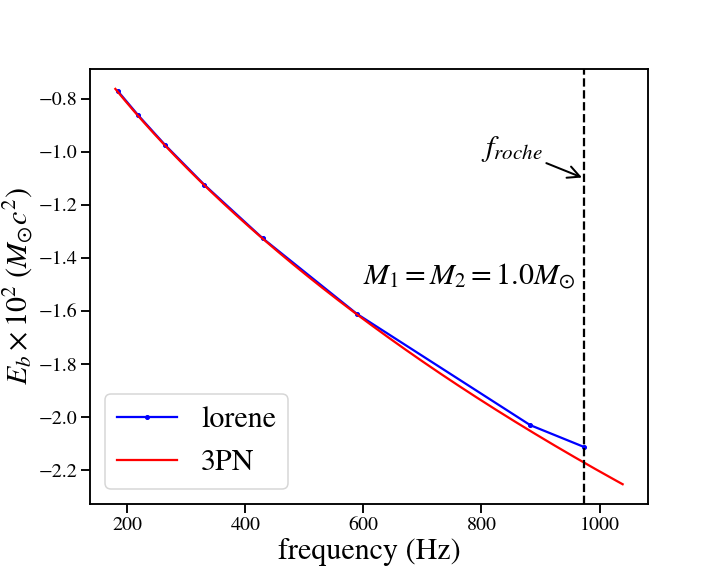}
    \caption{Gravitational binding energy $E_{b}$ for an equal mass binary neutron star system with $M_1 = M_2 = 1\ M_\odot$ as computed with LORENE (accounting for matter effects) and in the point-particle limit at 3PN. The APR equation of state is used in this calculation. The Roche lobe overflow frequency (vertical dashed line) is estimated as the frequency beyond which no equilibrium solution can be found.}
    \label{fig:be}
\end{figure}

We approximate the gravitational wave driven inspiral of the binaries as a sequence of quasi-circular orbits. We use the publicly available library LORENE \cite{lorene_url} to create binary equilibrium configurations within the extended conformal thin sandwich formalism \cite{Gourgoulhon:2007ue}. Neutron star matter is taken to be irrotational, as this well represents the late stages of the binary evolution as studied in \cite{Bildsten:1992my, Kochanek:1992wk}. The resultant field equations are then solved using a multi-domain spectral approach, as implemented in LORENE \cite{Bonazzola:1998qx}. We refer to \cite{Gourgoulhon:2000nn} for details on the mathematical formalism and implementation, where various tests of the method in different regimes are also presented.
\begin{figure*}[t]
    \advance\leftskip-0.5cm
    \includegraphics[width=18cm]{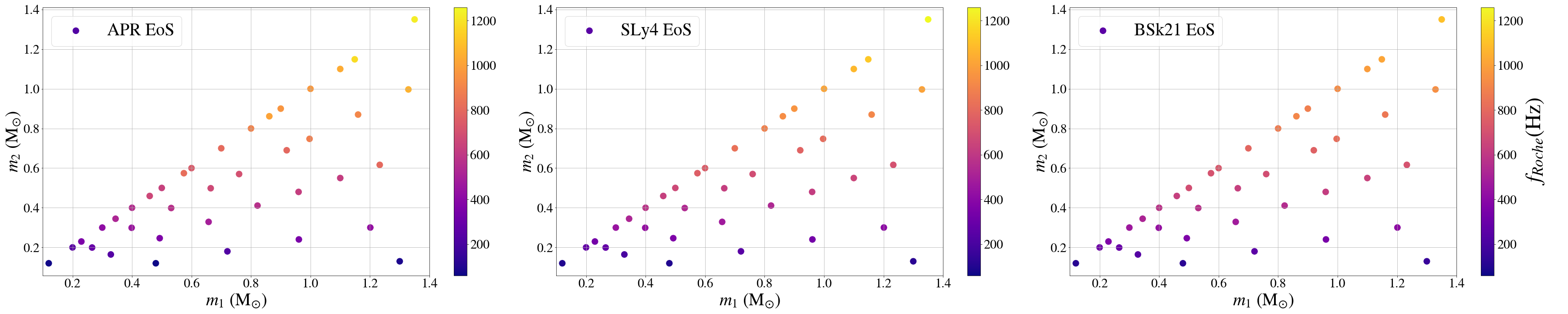}
    \caption{Gravitational wave frequency at Roche lobe overflow for binaries with different masses $m_1 \geq m_2$ and three equations of state: APR, SLy4, and BSk21. Each point represents a binary neutron star system, and is colored according to the gravitational wave frequency at the point of Roche lobe overflow.}
    \label{fig:m1m2fgw}
\end{figure*}  
For our calculations, we consider five domains for each neutron star, with one domain inside the star and the remaining 4 to cover the remaining region around them and match the boundary of the neutron star. The number of collocation points are $N_{r}\times N_{\theta}\times N_{\phi}=25\times17\times16$, where $N_{r}$, $N_{\theta}$, $N_{\phi}$ denote the number of points in radial, polar and azimuthal directions respectively.

Following \cite{Bejger:2004zx}, we monitor the gravitational binding energy of each binary $E_{b} = M_{\rm ADM} - (M_1 + M_2)$ as a function of the dominant ($\ell=2,m=2$) gravitational wave frequency $f_{\rm GW}$, where $M_{\rm ADM}$, the Arnowitt-Deser-Misner mass, is defined as the total mass-energy content in a hypersurface (see e.g. Eq.~64 of Ref~\cite{PhysRevD.63.064029}). We find good agreement between the numerical results and the 3PN point-particle limit \cite{Blanchet:2001id} at low frequency (large separations). As we increase the frequency (reduce the separation), we observe that the two start to deviate. This is an indication that matter effects become progressively more important and that the system is approaching Roche overflow \cite{Bejger:2004zx}. A representative example is shown in Fig.~(\ref{fig:be}). At even higher frequencies (smaller separations), no quasi-equilibrium configurations can be found using LORENE. We take the highest gravitational frequency for which a quasi-equilibrium configuration can be found as a conservative estimate for $f_{\rm Roche}$. To validate this choice, we perform full numerical relativity simulations of 3 configurations using the LORENE configuration with $f_{\rm GW} = f_{\rm Roche}$ as initial data and the \texttt{WhiskyTHC} code \cite{Radice:2012cu, Radice:2013hxh, Radice:2015nva} for the evolution. $f_{\rm Roche}$ should be thought of as a lower limit to the merger frequency. We find that the merger takes place only ${\sim}2{-}3$ gravitational wave cycles after this point~\cite{surendra}.

Using this procedure, we estimate $f_{\rm Roche}$ for various other mass configurations and the three equations of state considered in this study: APR, SLy4, and BSk21. In total, we have generated 117 binary sequences (39 for each equation of state). Our results are collected in Fig.~(\ref{fig:m1m2fgw}). The observed trends can be understood as an interplay between the compactness of the stars, the tidal fields of the companion, and the mass ratios. When the mass increases, the neutron star becomes more compact and less susceptible to the companion's tidal field in a binary. The tidal field falls off with distance as $d^{-3}$. As a result, only at very small separations the companion's tidal field becomes strong enough to initiate Roche lobe overflow. When considering different mass ratios, the trend is that, as the asymmetry in binary masses increases, the lower mass star is tidally disrupted at a larger separation.  
As can be seen from Fig.~(\ref{fig:m1m2fgw}), the Roche lobe overflow frequencies typically lie within the sensitive band for the Advanced LIGO detectors, thus suggesting that it is important to consider the effect of lower termination frequencies of the inspiral signals in modelling the gravitational wave signals from these sources.

\section{Impact on Search Sensitivity} \label{section4}

The discussion in Sec.~\ref{section2} and Sec.~\ref{section3} motivates the need to incorporate the tidal deformabilities of the neutron stars and their physical merger frequencies, for accurate modelling of gravitational waveforms from subsolar-mass binary neutron star systems. In order to test how the differences in the binary neutron star waveforms from their binary black hole counterparts affect the search sensitivity, we simulate a set of realistic binary neutron signals, and determine the ability to recover these waveforms for the Advanced LIGO detectors. The tidal deformabilities  of the neutron stars are derived from the APR, SLy4 and BSk21 equations of state, and their merger frequencies are calculated by using a two-dimensional interpolating function which estimates the Roche Lobe overflow frequency of the system as a function of component masses $m_1$ and $m_2$, using the data from the simulations described in Sec.~\ref{section3}.\\
Gravitational waves from compact binary coalescences are conventionally searched for by matched filtering against pre-constructed template banks \cite{Usman:2015kfa,Nitz:2018rgo,DalCanton:2020vpm,Adams:2015ulm,Messick:2016aqy,Sachdev:2019vvd,Hanna:2019ezx,Venumadhav:2019lyq,Venumadhav:2019tad,Olsen:2022pin,Chu:2020pjv}, which are families of templates generated using a parameterization that is believed to efficiently describe the astrophysical parameter space of interest. We used the Advanced LIGO zero-detuned high power noise curve \cite{KAGRA:2013rdx} and a lower frequency cutoff of 15 Hz for constructing the template bank, as well as for fitting factor computations. The aim of this paper is to explore how well the conventional binary black hole template banks used in LIGO searches would recover an incoming gravitational wave signal from subsolar-mass binary neutron star mergers. We restrict our analysis to non-spinning binary neutron star systems. In principle, waveform models for rotating neutron stars can be considered, as has been discussed in \cite{Stergioulas:2003yp}, but a detailed discussion of spin is beyond the scope of this work.  \\
In order to quantify how well a binary black hole template bank can identify a gravitational wave signal from a subsolar-mass binary neutron star system, we perform a set of template bank simulations and evaluate the fitting factors for each of the injected signals. For a given signal waveform $\tilde{h}_s$, the fitting factor is the maximum fraction of signal-to-noise ratio recovered when it is filtered against the template from the bank that generates the highest value of match for it \cite{PhysRevD.52.605,Brown:2012qf,Brown:2012nn,Huerta:2013qb}. It serves as a measure of the efficiency of the template bank in searching for gravitational wave sources characterized by the same parameters as those used to construct the template bank.
\begin{equation}
   \label{fittingfactor}
    FF = \underset{\tilde{h}_b \in \textrm{bank}}{\max} \mathcal{M}(\tilde{h}_b,\tilde{h}_s)
\end{equation}
where the match, $\mathcal{M}$, is the noise-weighted overlap between two normalized waveforms, maximized over the extrinsic parameters of the system \cite{Owen:1998dk}, which in this case, are the time $t_C$ and the phase $\phi_C$ of coalescence.

\begin{equation}
    \mathcal{M} (\tilde{h}_b(\Lambda),\tilde{h}_s(\lambda;t_c,\phi_c)) = \underset{t_C,\phi_C}{\max} \langle \tilde{h}_b(\Lambda)|\tilde{h}_s(\lambda)e^{i (2 \pi f t_c-\phi_c)} \rangle  \label{match}
\end{equation}

\begin{figure}
\centering
\includegraphics[width=0.5\textwidth]{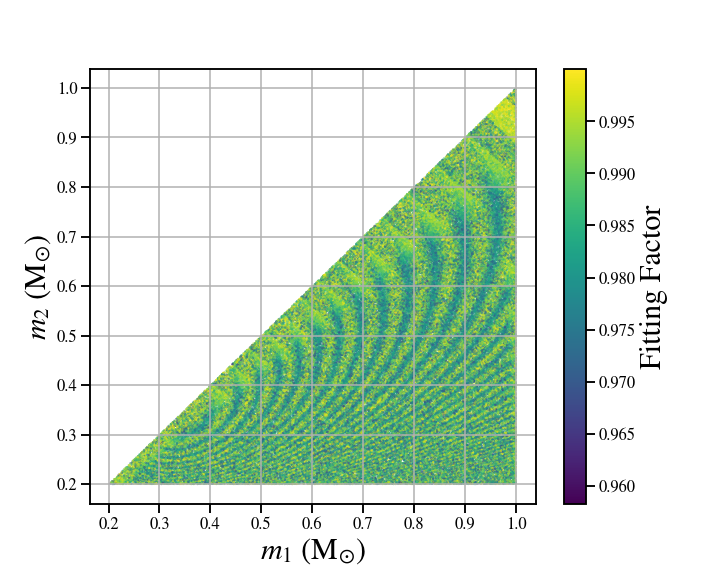}
\caption{The shaded region in the figure shows the fitting factor (Eqn.~(\ref{fittingfactor})) as a function of component masses $(m_1,m_2)$ of injected binary black hole signals. Fitting factors are highest for signals located closest to the templates, and lowest for those located equidistant from two nearest templates.  }
\label{bbh_banksims}
\end{figure}

To determine the loss in signal-to-noise ratio incurred on using a binary black hole template bank to search for subsolar-mass binary neutron star signals, we construct a template bank containing $1.7 \times 10^6$ templates which span the mass range $m_1,m_2 \in [0.2,1.0] \, \textrm{M}_{\odot}$. For details of template bank construction, we use Ref.~\cite{Owen:1998dk,Cokelaer:2007kx}. The templates are generated using TaylorF2 waveforms which follow the stationary phase approximation, accurate to 3.5 PN order. They have an upper cutoff frequency corresponding to that of gravitational wave emission from a point particle orbiting around a Schwarzschild black hole, in its innermost stable circular orbit (ISCO). This varies inversely with the total mass $M$ of the binary system, as
\begin{equation}
    f_{\textrm{ISCO}} = \frac{c^3}{6 \sqrt{6} \pi G M} \label{ISCO}
\end{equation}
The template bank is constructed to have a minimal match of 0.95, corresponding to a maximum allowed fractional loss in signal-to-noise ratio of $5\%$  due to the mismatch between the signal and the template. \\
To test the efficacy of the template bank, we perform a series of template bank simulations to evaluate the fitting factors for injected signals corresponding to binary black holes (used as the control setup), and binary neutron star inspirals with component masses lying in the range $m_1,m_2 \in [0.2,1.0] \; \rm M_{\odot}$, and having tidal deformabilities derived from the APR, SLy4 and BSk21 equations of state. To test the effect of the different frequency cutoffs, we perform one set of simulations where the inspiral waveforms extend all the way up to the Schwarzschild ISCO frequency, and another where they are terminated at the Roche Lobe overflow frequency, computed by using a two-dimensional interpolating function for the data generated from the simulations described in Sec.~\ref{section3}.\\
Iterating through the full template bank, containing 1.7 million templates, to find the best match template for a large number of injected signals, would cause a wasteful expenditure of a large amount of computational power and resources. We can reduce the number of match computations by comparing the values of the 0 PN chirp time, $\tau_0$, of the injected signal with those of the templates in the bank, and eliminating those templates that have a $\tau_0$ value that differs from that of the injection by more than 3 seconds. Here, the $\tau_0$ parameter depends on the chirp mass of the system as $\tau_0 = \frac{5}{256 \left( \pi f_L \right)^{8/3}} \mathcal{M}_C^{-5/3}$, where $f_L$ is the lower frequency cutoff~\cite{Babak:2006ty}. Since $\tau_0$ is a metric co-ordinate used in the construction of the template bank, the difference between the $\tau_0$ values of the signal and the template serves as a test of proximity of templates to the signal.\\
The results of the template bank simulations for the control setup, consisting of binary black hole signals, are shown in Fig.~(\ref{bbh_banksims}). The figure illustrates that the template bank recovers $100 \%$ of the injected subsolar-mass binary black hole signals with fitting factors greater than 0.95, as is expected. \\
\begin{table*}
\begin{ruledtabular}
\begin{tabular}{ccccc}
Equation of State and\\ Termination Frequency &Minimum Fitting Factor&Maximum Fitting Factor& $\%$ of signals having FF < minimal match \\ \hline
 APR (RLO)&0.6024&0.9823&63.06\\
 APR (ISCO)&0.5788 &0.9810&69.39\\
 SLy4 (RLO)&0.5780&0.9785&72.72\\
 SLy4 (ISCO)&0.5580&0.9761&78.63\\
 BSk21 (RLO)&0.5900 &0.9712&82.24 \\
 BSk21 (ISCO)&0.5764 &0.9690&87.12\\
\end{tabular}
\end{ruledtabular}
\caption{\label{tab:table1} A summary of the results of the template bank simulations, for different equations of state, and termination frequencies. ISCO stands for the frequency of gravitational wave emission by a point particle orbiting a Schwarzschild black hole, in its innermost stable circular orbit. RLO refers to the Roche Lobe overflow frequency, where one of the neutron stars is tidally disrupted by its companion and has reached the mass shedding limit.}
\end{table*}
For the subsolar-mass binary neutron star systems, the fitting factors are weakly dependent on the equation of state, but for a vast majority of the signals, they are lower than 0.95, which is the specified minimal match for the template bank. Table~\ref{tab:table1} summarizes the results of the template bank simulations for the binary neutron star signals. For the softer equations of state, the tidal deformabilities are low, and so the percentage of signals having fitting factors lower than 0.95 is less than that for the stiffer ones. However, as can be seen from the third column of the table, a majority of the injected signals can not detected by using the binary black hole template bank, with the percentage of signals having fitting factors lower than the minimal match ranging from $63\%$ for the APR equation of state, to $82\%$ for BSk21. \\
\begin{figure*}
\centering
\includegraphics[width=1.0\textwidth]{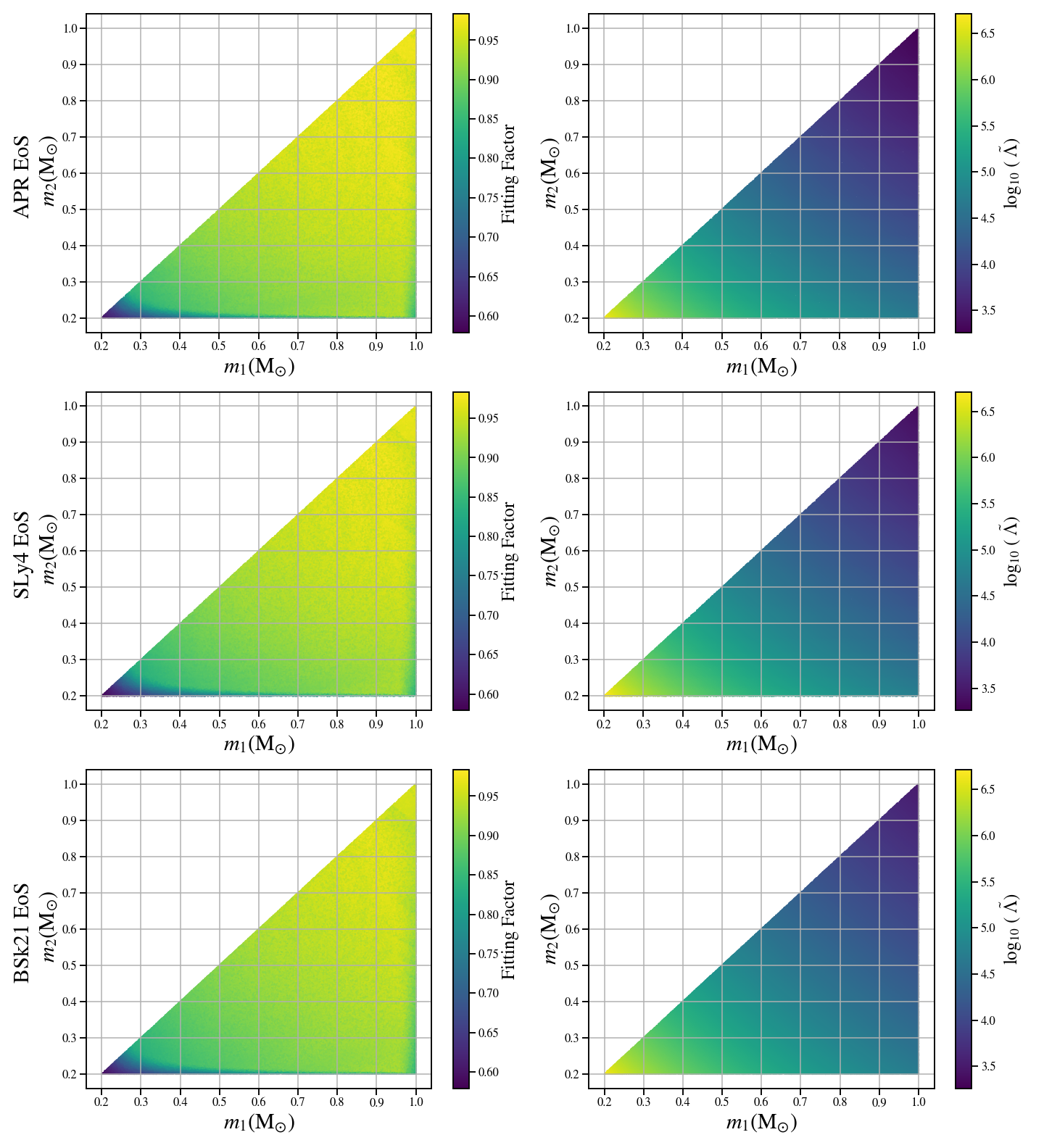}
\caption{Left panel shows fitting factors obtained on performing template bank simulations, for injected binary neutron star signals with the component neutron stars having tidal deformabilities derived from the APR, SLy4 and BSk21 equations of state, and terminated at their Roche Lobe overflow frequencies, plotted as a function of the primary mass $m_1$ and secondary mass $m_2$. Right panel shows a plot of effective tidal deformability of the binary neutron star systems, as a function of $m_1$ and $m_2$. The fitting factors are lower for systems having higher tidal deformabilities, indicating greater mismatch between the binary black hole templates and the injected binary neutron star signals.  }
\label{templatebanksimulations}
\end{figure*}
The fitting factors evaluated from the template bank simulations improve slightly when the waveforms are terminated at their Roche Lobe overflow frequency. However, even for the softest equation of state being considered here, i.e. the APR equation of state, the fitting factor obtained on matched-filtering an inspiral signal from a $(0.2,0.2) \; \textrm{M}_{\odot}$ binary neutron star system is $\sim 0.60$, corresponding to a $78.4 \%$ loss in sensitive volume of the detectors to these sources. Fig.~(\ref{templatebanksimulations}) shows the tidal deformabilities, as well as the fitting factors obtained from the template bank simulations, as a function of the component masses of the injected binary neutron star signals, for the three different equations of state. \\

\section{Constraint on Merger Rate} \label{section5}
\begin{figure*}[t]
\advance\leftskip-1cm
\includegraphics[width=1.1\textwidth]{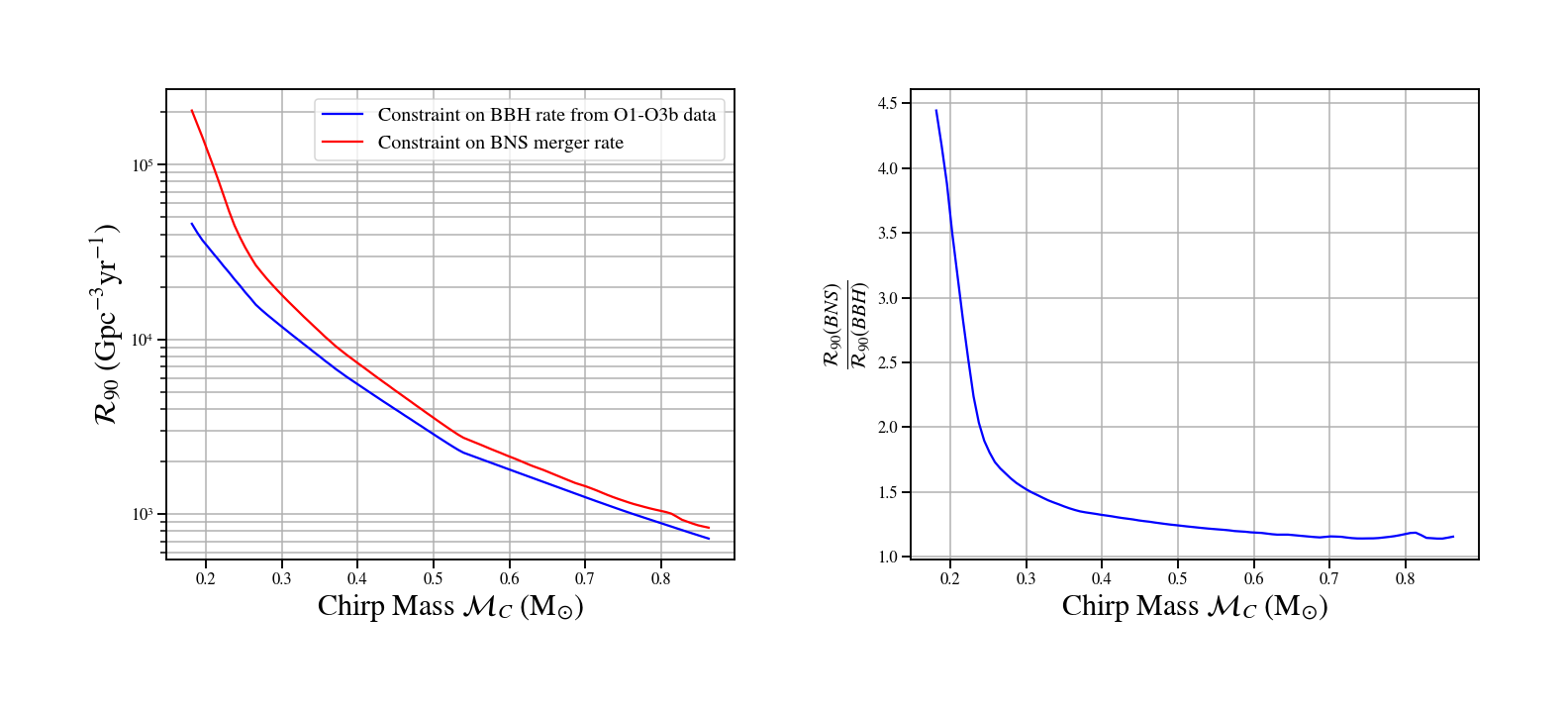}
\caption{Left panel: Blue curve shows the upper limit on merger rate for subsolar-mass binaries obtained in \cite{Nitz:2022ltl} for the O1 to O3b observing runs. Red curve shows an upper limit on the rate of subsolar-mass binary neutron star mergers, obtained using the sensitive volume reduced by a factor of $(\textrm{average fitting factor})^3$. Right panel: The ratio of upper limit on the rate of binary neutron star mergers to the upper limit on the rate of binary black hole mergers, plotted as a function of chirp mass of the binary system. }
\label{rate-upper-limit}
\end{figure*}

We use the loss in sensitivity computed in Sec.~\ref{section4}, and the results from the search for subsolar-mass binary black holes reported in \cite{Nitz:2022ltl}, to obtain a conservative upper limit on the merger rate of subsolar-mass binary neutron stars. This search used data collected by the Advanced LIGO and Advanced Virgo detectors during their first three observing runs, to place an upper limit on rate for subsolar-mass binaries, in the range $720 - \, 46000 \; \textrm{Gpc}^{-3} \textrm{yr}^{-1}$ for the chirp mass range $(0.17,0.87) \, \textrm{M}_{\odot}$. Using the loudest event statistic formalism, the $90 \%$ confidence interval on the chirp mass dependent upper limit of merger rate can be obtained as
\begin{equation}
    \label{merger_rate}
   \mathcal{R}_{90} = \frac{2.3}{\langle V T \rangle}
\end{equation}
where $V$ is the sensitive volume, and $T$ is the time analyzed~\cite{Biswas:2007ni}. \\
To estimate an upper limit on the merger rate of subsolar-mass binary neutron stars, we use the fractional loss in average sensitive distance, computed from the fitting factors. We compute the average fitting factors for the entire population by dividing into equally spaced chirp mass bins, and use them to compute the sensitive-volume of the detectors for binary neutron stars as $V_{BNS} = FF^3 \times V_{BBH}$. Substituting this in Eq.~(\ref{merger_rate}), we obtain a rate upper limit for subsolar-mass binary neutron stars as $[830-210,000] \, \textrm{Gpc}^{-3} \textrm{yr}^{-1}$ for the chirp mass range $(0.17,0.87) \, \textrm{M}_{\odot}$. \\
For the subsolar-mass binary black holes, the merger frequencies do not lie in the sensitive band of the LIGO-Virgo detectors, and the amplitude of the gravitational wave strain is a function of the chirp mass at leading order. As a result, the constraint on the merger rate for the primordial black hole search is not strongly dependent on the total mass of the sources or their mass ratios. Since we combine the results of the template bank simulations with the rate constraint from the primordial black hole search, and the fitting factors vary only slightly as a function of mass ratios, the rate constraint derived here is not sensitive to the mass ratios of the binary neutron star systems. It is also found to be largely independent of the equation of state model being used, since the fitting factors from the template bank simulations do not depend strongly on the equation of state. Fig.~(\ref{rate-upper-limit}) depicts the derived upper limit on the merger rate of subsolar-mass binary neutron stars. Our results indicate that a dedicated search for binary neutron star signals using a template bank that incorporates their tidal deformabilities and physical merger frequencies would be more sensitive to these sources than a search for subsolar-mass binary black holes. 

\section{Conclusions} \label{section6}

The analysis presented in this paper explores our ability to identify gravitational wave signals from inspiralling subsolar-mass binary neutron star systems, when a template bank search for compact binary coalescences involving low mass sources is carried out for the Advanced LIGO and Virgo detectors. As discussed in the text, existing template banks to search for binary neutron star signals treat neutron stars as effective black holes, since their tidal deformabilities and lower frequencies of merger are unaccounted for in the construction of the template banks. We find that ignoring the finite size effects of low mass neutron stars, such as their large tidal deformabilities and physical merger frequencies,  on the gravitational wave signals, can lead to a significant loss in signal-to-noise ratio, thus potentially missing signals from such sources. Using a binary black hole template bank to search for subsolar-mass candidate events in Advanced LIGO noise fails to recover a significant fraction of the injected signals with fitting factors above the specified minimal match for the template bank. This effect is amplified with decreasing chirp mass of the system. At the lower end of the mass range considered here, the loss in sensitive volume of the detectors is as high as $78.4 \%$, thus necessitating the use of refined template banks which are specifically designed to search for binary neutron stars~\cite{Harry:2021hls} in order to effectively recover gravitational wave signals involving subsolar-mass neutron stars. \\
Using the results of the search for subsolar-mass binaries for the O3b observing run of the LIGO-Virgo detectors~\cite{Nitz:2022ltl}, and accounting for the reduced detection efficiency for low mass neutron stars, we obtain a chirp mass dependent upper limit on the merger rate, lying in the range $[830-210,000] \, \textrm{Gpc}^{-3} \textrm{yr}^{-1}$. This constraint can be improved by implementing a search using a template bank that incorporates the tidal deformabilities and physical merger frequencies for binary neutron star signals, which we will explore in a subsequent paper. The cost of performing a search for subsolar-mass neutron stars would depend, to a large extent, on that of generating a template bank that spans the parameter space of interest, i.e. $m1,m2 \in [0.2,1] \mathrm{M}_{\odot}$ and their respective tidal deformabilities lying in the range allowed by plausible equations of state describing dense nuclear matter. Template banks incorporating tidal effects can be generated using a stochastic template placement algorithm, as discussed in \cite{Harry:2021hls}. Including the tidal deformability is expected to increase the number of templates beyond 1.7 million, which is the size of the bank spanning the $m1,m2$ parameter space for binary black hole systems. However, imposing physical constraints on the allowed values of tidal deformability, i.e. restricting to a region bounded by plausible equations of state on the soft and stiff end, can reduce the number of templates required to some extent. The ability to test the validity of proposed formation mechanisms for neutron stars and further constrain the nuclear equation of state, through the detection of gravitational wave signals involving low-mass neutron stars, makes it a useful scientific goal to consider a directed search for such events in the future observing runs of the LIGO-Virgo-KAGRA network.\\
Supporting data for this manuscript is available at \href{https://github.com/sugwg/sub-solar-ns-detectability}{https://github.com/sugwg/sub-solar-ns-detectability}.
\begin{acknowledgments}
The authors thank Alexander Nitz for helpful discussion regarding subsolar-mass binary black hole searches in LIGO and Virgo and making the data required to make Figure 7 available. DR acknowledges funding from the U.S. Department of Energy, Office of Science, Division of Nuclear Physics under Award Number(s) DE-SC0021177 and from the National Science Foundation under Grants No. PHY-2011725, PHY-2020275, PHY-2116686, and AST-2108467. DAB, AB acknowledge funding from the National Science Foundation under Grant No. PHY-2011655. DAB also thanks National Science Foundation Grant No. award PHY-2116686. EL acknowledges funding from the National Science Foundation under Grant No. AST-1559694. JP acknowledges funding from the U.S. Department of Energy Office of Science, Office of Nuclear Physics under Award Number DE-FG02-92ER40750. CJH and BR acknowledge funding from the U.S. Department of Energy Office of Science, Office of Nuclear Physics under Award Number DE-FG02-87ER40365. DAB and CJH thank the Kavli Institute for Theoretical Physics (KITP) for hospitality during the initial development of this work. KITP is supported in part by the National Science Foundation under Grant No. NSF PHY-1748958. This research was supported through computational resources provided by Syracuse University. 
\end{acknowledgments}

\let\itshape\upshape
\bibliographystyle{apsrev4-1}
\typeout{}
\bibliography{references} 

\end{document}